\documentclass[a4paper,fleqn,usenatbib]{mnras}

\usepackage{mathptmx}
\usepackage{txfonts}

\usepackage[T1]{fontenc}
\usepackage{ae,aecompl}

\usepackage{graphicx}	
\usepackage{amssymb}	

\title[Alignment of Radio Galaxies]{Alignments of Radio Galaxies in Deep Radio Imaging of ELAIS N1}

\author[A.\ R.\ Taylor and P.\ Jagannathan ]{
A.\ R.\ Taylor,$^{1,2}$\thanks{E-mail: russ@ast.uct.ac.za}
P. Jagannathan $^{1,3}$
\\
$^{1}$Department of Astronomy University of Cape Town, Rondebosch, South Africa \\ 
$^{2}$Department of Physic and Astronomy, University of the Western Cape, Street Address, Cape Town Postal Code, South Africa\\
$^{3}$National Radio Astronomy Observatory, Socorro, New Mexico, USA
}

\date{Accepted XXX. Received YYY; in original form ZZZ}

\pubyear{2015}

\begin{document}
\label{firstpage}
\pagerange{\pageref{firstpage}--\pageref{lastpage}}
\maketitle

\begin{abstract}
We present a study of the distribution of radio jet position angles of radio galaxies over an area of 1 square degree in the ELAIS N1 field.  ELAIS N1 was observed with the Giant Metrewave Radio Telescope at 612 MHz to an rms noise level of
10 $\mu$Jy and angular resolution of $6''\times 5''$.  The image contains 65 resolved radio galaxy jets.
The spatial distribution reveals a prominent alignment of jet position angles along a ``filament'' of about 1$^{\circ}$.
We examine the possibility that the apparent alignment arises from an underlying random distribution and find that the probability of 
chance alignment is less than 0.1\%. 
An angular covariance analysis of the data indicates the presence of spatially coherence in position angles
on scales $>0.5^{\circ}$.
This angular scales translates to a co-moving scale of $>20$\,Mpc at a redshift of 1. 
 The implied alignment of the spin axes of massive black holes that give rise the radio jets 
suggest the presence of large-scale spatial coherence in angular momentum. 
Our results reinforce prior evidence for large-scale spatial alignments of
quasar optical polarisation position angles. 

\end{abstract}

\begin{keywords}
radio continuum: galaxies -- galaxies: jets -- galaxies: statistics -- 
cosmology: large-scale structure of the universe
\end{keywords}



\section{Introduction}
There have been several observational studies to detect deviations
from isotropy in the orientation of galaxies following 
approaches first devised by \cite{1975AJ80477H}.
If detected, the presence of alignments and certain preferred orientations
can shed light on the origin and evolution of the galaxies and the
relation to large scale structure. 
Alignment might arise from the large scale environmental influences during
galaxy formation or evolution. Cosmic magnetic fields have been shown
to be present on scales of galaxy clusters and larger \cite{Ratra1992ApJ391L1R}. Effects of
seed magnetic fields from inflation \cite{Ratra1992ApJ391L1R},
axionic fields post inflation, and cosmic strings are possible
candidates that could effect an alignment in galaxies even on scales
larger than galaxy clusters.

Evidence for very large scale coherent orientations of quasar polarisation
vectors was provided by \cite{Hutsemeker1998}, who 
found large-scale coherence in polarisation position angles 
of galaxies around the Northern and Southern galactic
poles from a sample of 170 quasars. 
Initial Kuiper tests to show deviations from uniformity proved inconclusive
on the complete sample. However a statistical analysis based
on nearest neighbours showed evidence for alignments in quasar sub-samples
particularly around the north and south galactic poles with 99.99\%
significance level. Follow up observations by \cite{Hutsemeker2001A&A367381H} indicated
that the alignments in quasars polarisation vectors extends to 
comoving scales of $1000h^{-1}$Mpc  and that alignment
of quasar polarisation angles may be correlated with the large scale structure
it is embedded within.

The tendency for the axes of double-lobed
radio quasars to be aligned with the electric vectors of optical polarisation
in the active galactic nuclei is well known (\cite{1979ApJ227L55S}). 
{\cite{rusk1990} show that there is a correlation between the optical
polarisation and the structural axis of a galaxy, and \cite{battye2009radio} 
show that radio jets are aligned with the optical minor axis.
AGN jets position angles are thus a proxy for the direction of the major 
axis of the host galaxy.
{Measurements derived from the total intensity radio emission of AGN jets has the advantage of not
being affected by propagation effects such as scattering, extinction or
Faraday Rotation, which may be an issue for optical and polarimetric studies. 

{We have undertaken a deep radio imaging survey of the ELAIS-N1 region using 
the Giant Metrewave Radio Telescope (GMRT) as part of a larger, low-frequency study 
of the properties of faint radio sources at $\mu$Jy flux densities.
Such sensitive radio images, capable of detecting radio sources
in the $\mu$Jy regime and covering scales of a degree or more, offer a first
opportunity to use radio jets to explore alignments or radio galaxies on physical scales 
below a few 100 Mpc.
An angle of 1 degree spans a comoving distance of 40$h^{-1}$\,Mpc at redshift $z=1$
or 64$h^{-1}$\,Mpc at redshift $z=2$.  In this paper we report an analysis 
of the spatial correlation of radio galaxy jet position angles in the deep GMRT image of the  
ELAIS-N1 field,  and the detection of an alignment of the jets of a sub-sample of radio galaxies 
over angular scales of  0.5 degrees. 

\begin{figure*}
 \includegraphics[width=1.6\columnwidth]{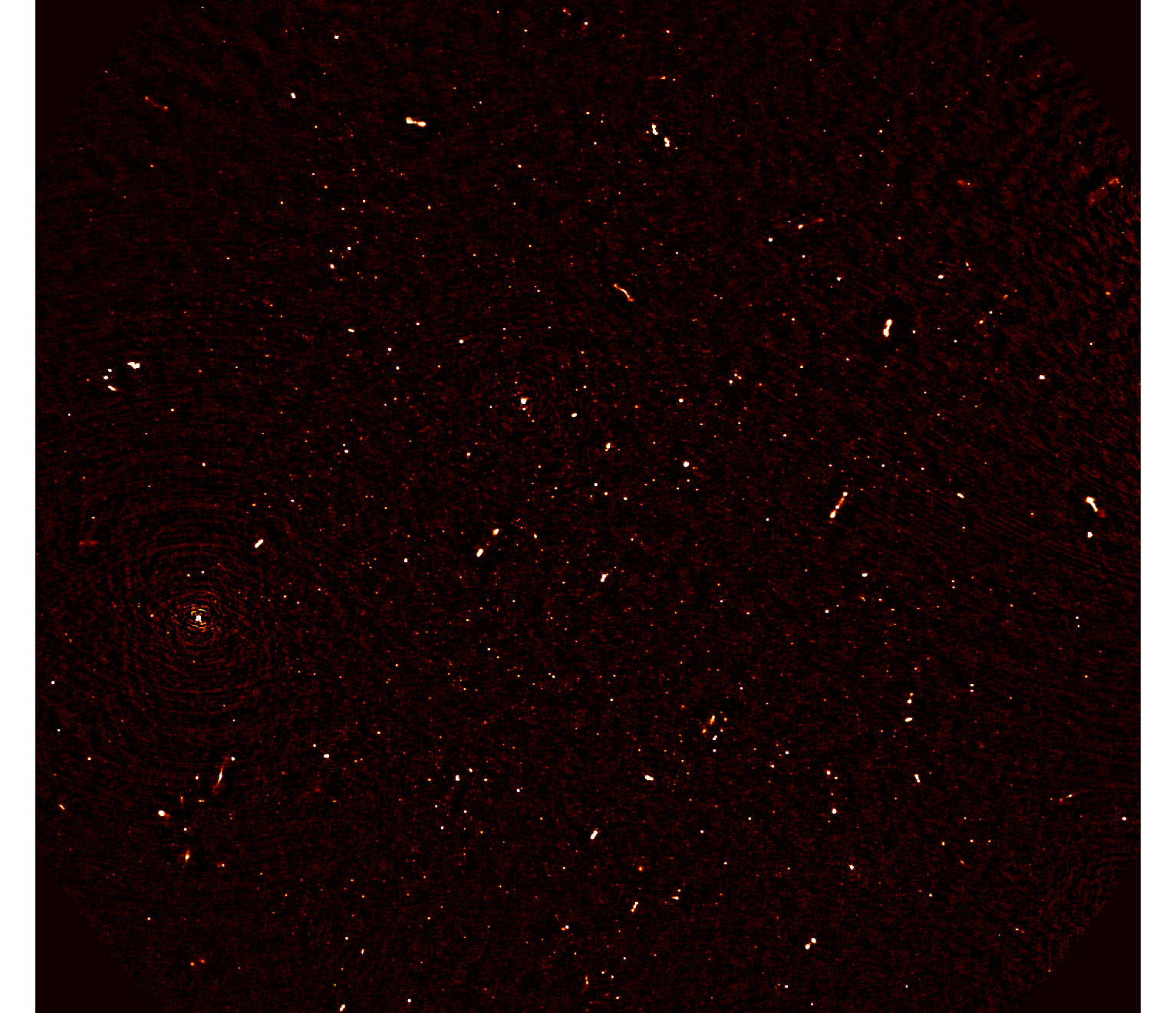}
 \caption{The GMRT 612 MHz total intensity image if ELAIS N1. The image covers 1.2 $\times$ 1.2 degrees
 with an RMS noise level of 10.3\,$\mu$Jy.  Several tens of double-lobed radio galaxies dominate the bright source
 population  in the image. The fainter sources are primarily compact, point-like objects that are unresolved at the 6$''$ resolution of this
 image, and a mostly radio emission from star-forming galaxies. }
 \label{fig:GMRTimage}
\end{figure*}

\section{Observations}

We carried out deep imaging observations of the ELAIS N1 field  with
the Giant Metrewave Radio Telescope (GMRT) in several observing
sessions from  2011 to 2013. 
An area of 1.2 sq degrees of ELAIS N1 was covered by a mosaic of 
7 pointings arranged in a hexagon pattern centred on 
$\alpha$ = 16$^{\rm h}$ 10$^{\rm m}$ 30$^{\rm s}$,  
$\delta$ = 54$^{\rm o}$ 35 00$''$.  

To reduce the noise and mitigate the effect of side-lobes from
off-axis gain errors in the central regions of the mosaic, 
the separation of the  pointings on the sky was closely spaced at 
16$'$, or 38\% of the FWHM of the GMRT primary beams at 612 MHz.
Each pointing was observed for
approximately 30 hours in three 10-hour sessions. 
Data were taken in 256 spectral channels in four
polarisation states covering a 32 MHz bandwidth centred at 612 MHz. 
The flux scale, bandpass and absolute polarisation position angle 
calibration was  secured by observations of 3C286 twice in each 
observing session.  
Time dependent complex gains and on-axis polarisation leakage corrections 
were measured by frequent observations of J1549+506.    

The visibility data were calibrated, deconvolved (CLEANed), self-calibrated, 
imaged, and the individual fields mosaicked together using the 
CASA processing software.  
The central 1.1 sq degrees of the GMRT mosaic image are shown in 
Figure~\ref{fig:GMRTimage}.
The rms noise in this mosaic is 10.3\,$\mu$Jy/beam before primary beam correction.   
The angular resolution (FWHM of the synthesised beam) is 6.1$'' \times$ 5.1$''$.

\section{Alignments of Radio Sources}

\subsection{Source Finding }

Within the region defined by the half-power point of the 6 outer pointings, there
are 2800 sources above a flux density of 50\,$\mu$Jy.     
A vast majority of the sources are faint, compact sources that are unresolved by the 6$''$ synthesised beam. 
However, the brighter sources in the image are dominated by objects that exhibit classical FRI or FRII elongated radio galaxy jet morphologies.
These sources were identified on the image visually and the coordinates of the two end points of the jets for each source were measured from the images.
This processed yield a set of 64 radio galaxies that constitutes a complete sample of jet sources down to the resolution limit of the observations. 
This sample forms the basis of our further analysis.

\subsection{The Distribution of Position Angles}
\begin{figure}
 \includegraphics[width=\columnwidth]{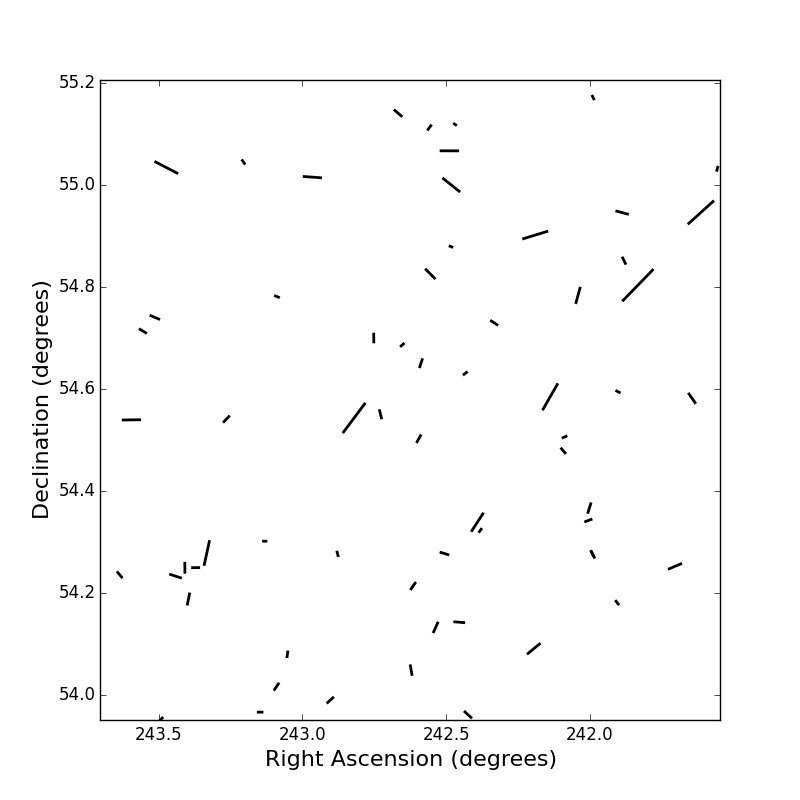}
 \caption{This stick diagram shows the direction and lengths of radio jets 
 at the positions of radio galaxies in the ELAIS-N1 612 MHz radio image. 
To enhance visibility, the lengths of the jets have been expanded by a 
factor of two.}\label{fig:stick}
\end{figure}

Figure~\ref{fig:stick} shows the positions, directions and lengths of
the 64 radio galaxy jet in the field.  To render the objects
more easily viewable, the size of the vector for each of the jets is plotted 
at twice their true length.  
The plot shows a number of large jets with similar position angle of
about -40$^{\circ}$ running along a locus starting at the lower middle of the 
image and extending to upper right.  
To assess whether such apparent alignments arises by chance, or whether 
they reflect real spatial correlation in the directions of radio jets 
we explore the statistical likelihood of the alignments given an
underlying random distribution of jet directions.

For analysis involving position angles on a sphere, standard estimators
such as the mean need to be redefined. We can clearly see that $0^{\circ}$
and $360^{\circ}$ are the same. Similarly the mean angle of
two sources with position angles $2^{\circ}$ and $358^{\circ}$ is
not $180^{\circ}$. \cite{fisher1993} provides a comprehensive collection
of methods of dealing with spherical statistics.

We can consider the position angles to be unit vectors on a circle
in which case each angle, $\phi$, can be thought of as a Cartesian
point on the edge of the circle with coordinates $(\cos\phi, \sin\phi)$.
To compute the arithmetic mean we could simply take the arithmetic
means of the Cartesian points and then convert them back into polar
form to obtain the mean angular measure. 
Given angles $\phi_{1},\phi_{2},...,\phi_{n}$
the mean angle is given by 
\begin{equation}
\bar{\phi}=arg(\frac{1}{n}\sum_{j=1}^{n}e^{i\phi_{j}})
\label{mean_eq}
\end{equation}
and the mean resultant length by
\begin{equation}
\bar{r} = \sqrt{ \overline{\cos \phi}^2 + \overline{\sin \phi}^2}
\end{equation}
where $\overline{\cos \phi}$ and $\overline{\sin \phi}$ are the average values of $\cos \phi_i$ and $\sin \phi_i$.

Like polarisation angle, the position angle of radio jet sources is symmetric to a change by $\pi$.
Values are thus defined in the range$[-\pi/2,\pi/2]$ degrees. 
For a parent random distribution of position angles, the probability distribution
of the positions angles on the circle should be a uniform within this range,
with every angle equally likely, implying that the probability
distribution function is 

\begin{equation}
f_{UC}=\frac{d\phi}{\pi} .
\end{equation}

\begin{figure}
 \includegraphics[width=1.05\columnwidth]{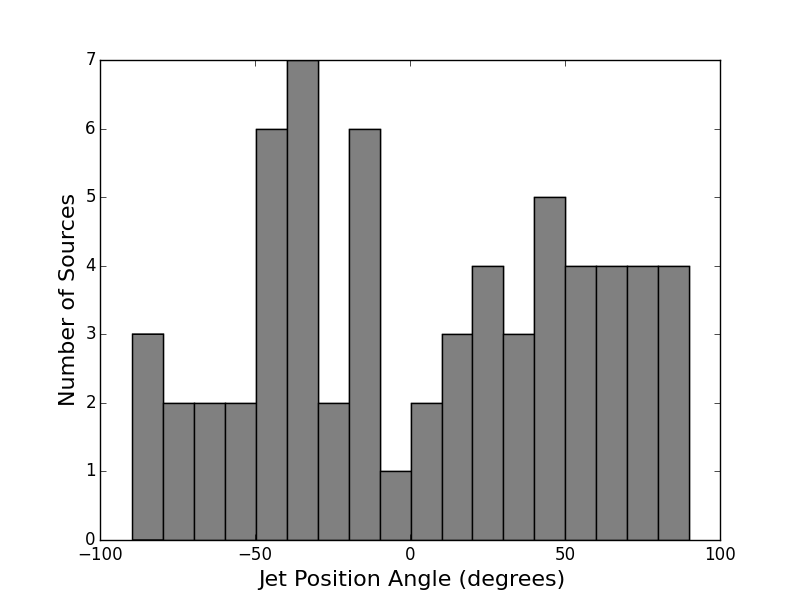} 
 \caption{The distribution of position angles of the radio galaxy jets shown
 in Figure~\ref{fig:stick}.}
 \label{fig:PDF-PA}
\end{figure}

\begin{figure}
\begin{centering}
\includegraphics[width=1.05\columnwidth]{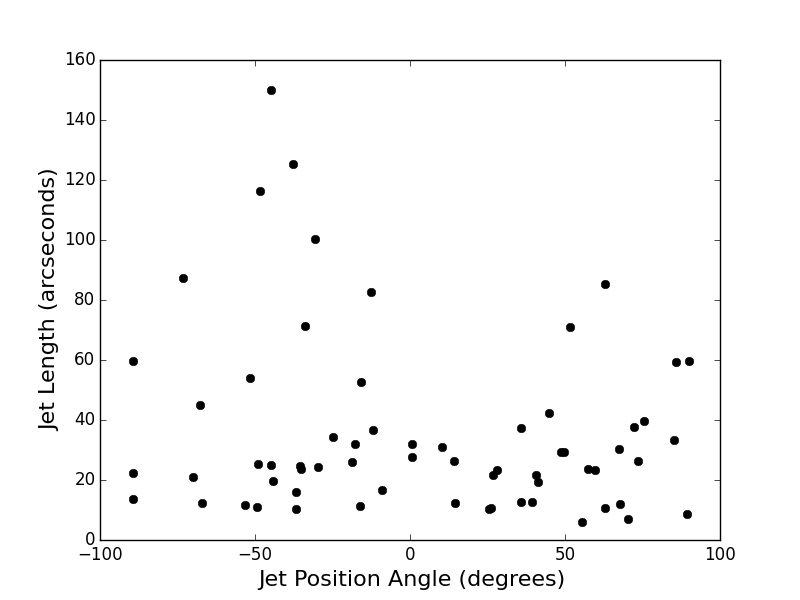}
\end{centering}
\caption[PA versus Length] {The length of the 64 radio jets plotted against
jet position angle. The longest jets are preferentially present in the 
excess of object with polarisation angle $\sim -40^{\circ}$.
} 
\label{fig:lengths}
\end{figure}

Figure \ref{fig:PDF-PA} shows the distribution of
positions angles of the jets.  For 64 sources the expectation value per bin in the 
case of a uniform distribution is the same and equal to 3.55.
For position angles greater than about 20$^{\circ}$, the
distribution appears consistent with uniform value.  
However, on the other side of the distribution there is a
strong peak from position angle, $\phi$, -50$^{\circ}$ to -30$^{\circ}$,
and a paucity of objects around 0$^{\circ}$ and with position 
angles less than -50$^{\circ}$.

Several of the objects in the excess between $-50^{\circ} < \phi < -30^{\circ}$ are
apparent in Figure~\ref{fig:stick} as the larger jets.  
This effect is illustrated in Figure~\ref{fig:lengths} which plots position angle
versus jet length.  The short jets (less than about 20$''$) appear to have a more uniform 
position angle distribution.  The longer jets appear clustered with a peak around
$\phi \sim -40^{\circ}$.

A number of available non-parametric tests are applicable to determine the probability of deviation of an observed distribution given the null hypothesis that the position angles of the jets are uniformly (randomly) distributed. 
For the analysis of our data we used the CircStat \citep{jammalamadaka2001topics} \& Circular \citep{circular} statistical package in R.

{The Rayleigh's test of uniformity measures the significance of the mean resultant length, $\bar{r}$, compared to a uniform distribution with a
statistically null mean resultant length. Since polarisation angle distributions are defined only on the half circle interval,  to populate the full circle for the Rayleigh test we used the angle doubling technique to test against the null hypothesis of uniformity of the distribution of $2\phi$.  The Rayleigh tests assumes the data are unimodal, i.e.\ there is at most one significant clustering of points around the circle. 
The mean resultant value of the observed angle distribution, $\bar{r}$, is 0.637, and the probability of this result from a 
parent random distribution is $<0.001$.

We also carried out two additional tests that examine the statistics of the differences between the observed and null hypothesis distributions.
The Kuiper test gives a probability of random of $<0.01$.  The Watson U$^2$ test which is similar to the Kuiper test but uses the mean square deviations rather than the maximum deviations of the difference distributions gives a similar result.  The results of all three tests are listed in Table~\ref{tab:tests}.

\subsection{Spatial Covariance of Position Angles}

\begin{table}
\begin{centering}
\begin{tabular}{|c|c|c|c|}
\hline  
Test & Statistic & P-value & Uniformity\tabularnewline
\hline 
\hline
Rayleigh's Test & $\bar{r}=0.6376$ & <0.001 & Non Uniform\tabularnewline
\hline
Kuiper's Test & 4.1066 & < 0.01 & Non Uniform\tabularnewline
\hline 
Watson's $U^2$ Test & 1.3819 & < 0.01 & Non Uniform\tabularnewline
\hline 
\end{tabular}
\par\end{centering}
\caption{Test for deviations from the Uniform Distribution (Null hypothesis)\label{tab:tests}}
\end{table}

Given that distribution of jet position angles is not random, it is of 
interest to explore the angular scale of the non-uniformity. One approach
is to calculate the spatial semi-variance, or the variogram. In spatial statistics the variogram is a graphic describing the degree of spatial dependence of a vector field. The semi-variance function is defined as 

\begin{equation}
\gamma(d)=\frac{1}{2m(d)}\sum_{j=1}^{m(d)}[z(x_{j})-z(x_{j}+d)]^{2}
\label{eqn:semivariance}
\end{equation}

Here the sum is over $m(d)$ pairs of points separated by a distance $d$ from each other,and $z$ is the variable being measured at vector location $x_{j}$. In the case of stationary and isotropic spatial process the semi-variance reduces to a spatial covariance function. 

If the distribution of the position angles of the jets is random then the semi-variance should be constant over all angular scales.  
For a random distribution of angles the distribution of the differences of 
any two angles over a range symmetric about zero is a triangular distribution with a mean of zero. 
We carried out a simple Monte Carlo simulation by drawing a million pairs of random position angles over the range [-$\pi/2, \pi/2$], and measuring the variance of the distribution of differences. 
The expectation value of $\gamma(d)$ for random angles is 0.82.  
This value is shown as the dotted line in the variogram for the observed jet
position angles in Figure~\ref{fig:variogram}. 
The calculations of the semi-variance were carried out in R using the 
geoR tools \citep{geoR, geostat}.
The error bars on the data points were determined from the number of pairs 
used to calculate the semi-variance in each angular separation bin.
We carried out 10,000 simulations for each bin, in which we pulled $N(d)$
pairs of random angles, where $N(d)$ is to the number of observed pairs 
in the bin with separation $d$.
The error was taken as the standard deviation about the mean (0.82) of
the 10,000 semi-variance values. There error bars are thus the 
$1\sigma$ error expected for a random distribution with the same number of pairs. 

Angular scales that exhibit deviations from a value of 0.82 indicate a departure of the distribution of angle differences from random. 
Points that lie below the line indicate a smaller dispersion of angle differences compared to random, indicating an alignment of angles between sources.  
Points above the line have larger dispersion than expect from random, 
indicated a preference for large angular differences, or anti-alignment. 
One situation that gives rise to anti-alignment is the presence of two distinct 
populations of aligned objects separated by the angular scale in question.
The low point in Figure~\ref{fig:variogram} suggest an alignment of position angles on scales of about 0.5 to 0.8 degrees. This is consistent with the 
angular scale of the "filament" of aligned large jets.  The indication of 
a rise of the semi-variance above 0.82 at larger angular scales tentatively
suggests the presence of two areas of alignment separated by scales larger than
$1^{\circ}$.  However at this scale we reach the limits of the dimension of
the GMRT image.  Imaging of a larger region to similar sensitivity will be 
required to explore potential spatial correlations on larger scales.

\begin{figure}
\begin{centering}
\includegraphics[width=1.05\columnwidth]{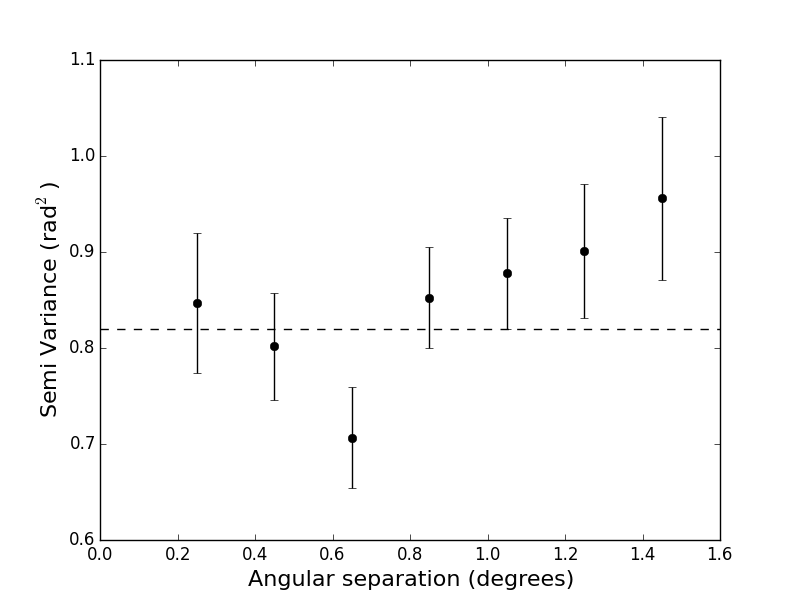}
\end{centering}
\caption[Variogram for the complete sample of AGN] {Semi-variance for the AGN sample
(Equation~\ref{eqn:semivariance}) as a function of angular separation
in degrees. The horizontal dashed line indicates the value expected for a random
parent distribution of position angles. 
} 
\label{fig:variogram}
\end{figure}

\section{Discussion and Conclusion}

The analysis of the probability and spatial distributions of radio galaxy jet 
position angles in ELAIS-N1 GMRT deep field indicates the presence of 
spatially correlated alignments on scales greater than about 0.5$^{\circ}$. 
ELAIS-N1 is one of the deep fields covered by the Spitzer Extragalactic Representative 
Volume Survey (SERVS) \citep{2012PASP..124.1135M}
photometric redshift catalogue.  However only 6 SERVS objects with redshifts 
were matched to the radio galaxies, and none in the major alignment of objects around
$\phi = -40^{\circ}$.  
We are thus unable to directly determine the redshift of the alignment feature.
{From photometric analysis of the UKIDSS data  \cite{2007MNRAS.379.1343S} identified  
fifteen cluster candidates within a 0.8 square degrees region of ELAIS N1.  Spectroscopic
follow-up of five candidates showed significant over densities for all
five candidates at  $z = 0.89$.  
\cite{2007MNRAS.379.1343S} infer the presence of a supercluster at this redshift that
extends over the entire field.
The aligned radio galaxy jets may arise from AGN hosted by giant elliptical galaxies associated 
with components of the supercluster.
 At redshift of $z = 0.9$ an angular scale of $0.5^{\circ} - 1.0^{\circ}$ 
corresponds to co-moving physics scale of $20h^{-1} - 40h^{-1}$\,Mpc.

From analysis of optical polarisation position angles of 19 quasars belonging to a quasar
group at $z \sim 1.3$,  \cite{2014A&A...572A..18H} presented
evidence of preferential alignment of the polarisation angles either parallel or
perpendicular to the large scale structure in which they are embedded. 
The statistical probability that their result is due to random orientations is or order 1\%.
Our result provides independent confirmation of large-scale spatial alignment of AGN axes,  with a
tracer that is independent of propagation effects along the line of site.

The direction of radio galaxy jets is determined by the direction of the 
angular momentum axis of the super massive black hole that drives the AGN
activity in the host galaxy.  The observed jet alignment thus implies an
alignment of the angular momentum axes of central black holes on scales of
several 10's of Mpc or greater. 
Such an alignment must arise at the time of formation and would imply spatially 
coherent angular momentum features of this scale embedded in the
local large-scale structure at early times. It would be interesting to compare
this implication with predictions of angular momentum structure from universe simulations.

Deep radio imaging, in combination with redshift surveys,
offers a powerful means for studies 
of the alignment of radio galaxies jets to high redshift and the relation of the angular distribution
of position angles to large scale structure. 
As shown in this study, at sensitivity of a few $\mu$Jy about 100 radio galaxy jets per square degree are 
detected, and the sampling of the population of classical radio galaxies 
is virtually complete. The fainter population is dominated by star-forming galaxies
and radio quiet AGN.  
Combining angular resolution of a few arc seconds with imaging areas of several square degrees
will allow the spatial distribution of position angles on physical scales of 100's of Mpc to 
high redshifts. 
Such imaging projects are in the planning stage for the Square Kilometre Array and
its precursor telescopes, the South African MeerKAT array and the Australian SKA Pathfinder 
(ASKAP).

\section*{Acknowledgements}
We thank the staff of the GMRT that made these observations possible. GMRT is run by the National Centre for Radio Astrophysics of the Tata Institute of Fundamental Research. PJ thanks the US National Radio Astronomy Observatory for providing support in the form of the Reber Doctoral Fellowship.

\bibliographystyle{mnras}
\bibliography{alignment}

\bsp	
\label{lastpage}
\end{document}